\documentclass[preprint,12pt]{elsarticle}
\usepackage[utf8]{inputenc}
\usepackage{url}
\usepackage{graphicx}
\usepackage{subfigure}
\graphicspath{{figs/}}
\usepackage{microtype}
\usepackage[section,above]{placeins}
\usepackage{textcomp}
\usepackage{textgreek}
 
\newtheorem{finding}{Finding} 

\usepackage[version=3]{mhchem}
\usepackage[detect-all]{siunitx}
\DeclareSIUnit{\Molar}{M}
\usepackage{xcolor}

\journal{Journal}

\begin{document}

\begin{frontmatter}

\title{On resistance switching and oscillations \\ in tubulin microtubule droplets}

\author[1]{Alessandro Chiolerio} 
\author[2]{Thomas C.\ Draper}
\author[2,3]{Richard Mayne}
\author[2]{Andrew Adamatzky}

\address[1]{Istituto Italiano di Tecnologia,
Center for Sustainable Future Technologies,   
Via Livorno 60, 10144 Torino, Italy}

\address[2]{
Unconventional Computing Laboratory,
University of the West of England,
Coldharbour Lane, Bristol, BS16 1QY, UK}

\address[3]{
Department of Applied Sciences,
University of the West of England,
Coldharbour Lane, Bristol, BS16 1QY, UK}


\begin{abstract}
We study electrical properties of Taxol-stabilised microtubule (MT) ensembles in a droplet of water. We demonstrate that the MT droplets act as electrical switches. Also, a stimulation of a MT droplet with a positive fast impulse causes oscillation of the droplet's resistance.  The findings will pave a way towards future designs of MT-based sensing and computing devices, including data storage and featuring liquid state.
\end{abstract}


\end{frontmatter}

\section{Introduction}

The protein tubulin is a key component of the eukaryotic cytoskeleton~\cite{lowe2001refined,cytoskeletonbook,Huber2013}. The networks of tubulin microtubules (MTs)~\cite{purich1984microtubule} are involved in cells' motility~\cite{fletcher2010cell,vicente2011cell,Golde2018}, intra-cellular transport~\cite{volkmann1999actin,schuh2011actin} and cell-level learning ~\cite{hameroff1988coherence, rasmussen1990computational, ludin1993neuronal, conrad1996cross, tuszynski1998dielectric, priel2006dendritic, debanne2004information, priel2010neural, jaeken2007new,dayhoff1994cytoskeletal}. In 1980 Hameroff and Rasmussen proposed that MT are responsible for a sub-cellular information processing~\cite{rasmussen1990computational,hameroff1990microtubule}. The framework of the sub-cellular computation  was further supported by Priel, Tuszynski and Cantiello  in their studies of information processing in actin-tubulin networks of neuron dendrites~\cite{priel2006dendritic}. The pathways towards development of cytoskeleton based computing devices~\cite{adamatzky2018towards,mayne2015cytoskeleton,mayne2019ca} have been so far mainly based on theoretical studies and computer modelling. This might be explained by the fact that experimentally-wise it might be difficult to address individual units of tubulin polymers to write and read information arbitrarily. Therefore, at the present it might be reasonable to focus on `tubulin electronics': designing of electronic components and circuits from MT networks with aim of their use in hybrid electro-mechanical analog computers. Another exciting framework is that of liquid robotics, where smart fluid systems could provide volume operation capabilities, being resilient to shape changes, impulse mechanical stresses, high pressures and radiation, and harsh environments in general~\cite{chiolerioquadrelli}. The liquid state provides means for holographically addressing the information~\cite{preston} and achieving much higher volume densities, also reducing routing dissipation and physical distance between bits of information~\cite{chioleriodraperadamatzky}. There is also the scope to further `armour' the droplets, by coating them with a hydrophobic powder, and thus form a `liquid marble'~\cite{Aussillous2001,Draper2019,Fullarton2018}.
Tuszynski and colleagues experimentally demonstrated memristive~\cite{tuszynski2016microtubule} and capacitive~\cite{kalra2019capacitive} electrical properties of a single MT. Priel and colleagues have experimentally shown, using patch-clamp technique, that MT exhibits electrical amplification and thus can act as bio-transistors~\cite{priel2006biopolymer}. Based on circumstantial evidences we can also claim that MT can act a variable resistors, because the conductivity of MT bundles can be controlled by stabilising and destabilising drugs~\cite{gharooni2019bioelectronics}. 

Advancing our ideas on computing with vesicles filled with reactive cargo~\cite{adamatzky2011vesicle,adamatzky2011computing,holley2011computational,mayne2015computing} we decided to check electrical properties of MT networks in a bulk volume, i.e.\ a droplet. The paper presents an experimental setup, Sect.~\ref{setup}, for interfacing with MT droplets, and findings on electrical switching and resistance oscillations measured after submitting the droplet to a fast pulse waveform, Sect.~\ref{results}. Applications of the findings are discussed in Sect.~\ref{discussion}.

\section{Experimental}
\label{setup}

Stabilised porcine brain microtubules (MTs) were supplied as a lyophilised powder. The MTs were reconstituted following the manufacture's instructions~\cite{tubulinedatasheet}, to yield the MTs. A buffer solution (\SI{10}{\milli\L}) was made using deionised water (DIW,  \SI{15}{\mega\ohm\centi\metre}), constituting 1,4-piper\-azine\-di\-ethane\-sulfonic acid (PIPES, \SI{15}{\milli\Molar}), \ce{MgCl2} (\SI{1}{\milli\Molar}), and adjusted to pH 7.0 using NaOH. Separately, paclitaxel (Taxol) was dissolved in anhydrous dimethyl sulfoxide (DMSO) (\SI{100}{\micro\L}, \SI{2}{\milli\Molar}). The Taxol solution was then added to the buffer solution, forming the `resuspension buffer' (RB). The RB (\SI{500}{\micro\L}) was then added to the lyophilised MT (\SI{500}{\micro\g}), followed by gentle stirring for 10 minutes, yielding a final concentration of \SI{1}{\milli\g\per\milli\L}. Lyophilised MTs and Taxol were stored at \SI{3}{\celsius}, and all reaction steps were conducted at room temperature. Lyophilised MTs (\#MT002, \SI{99}{\percent}) and Taxol (\#TXD01) were sourced from Cytoskeleton Inc; PIPES \& \ce{MgCl2} were from Sigma-Aldrich; and NaOH was from Fisher Scientific. All reagents were used as received.

\begin{figure}[!tbp]
    \centering
    \subfigure[]{\includegraphics[scale=0.3]{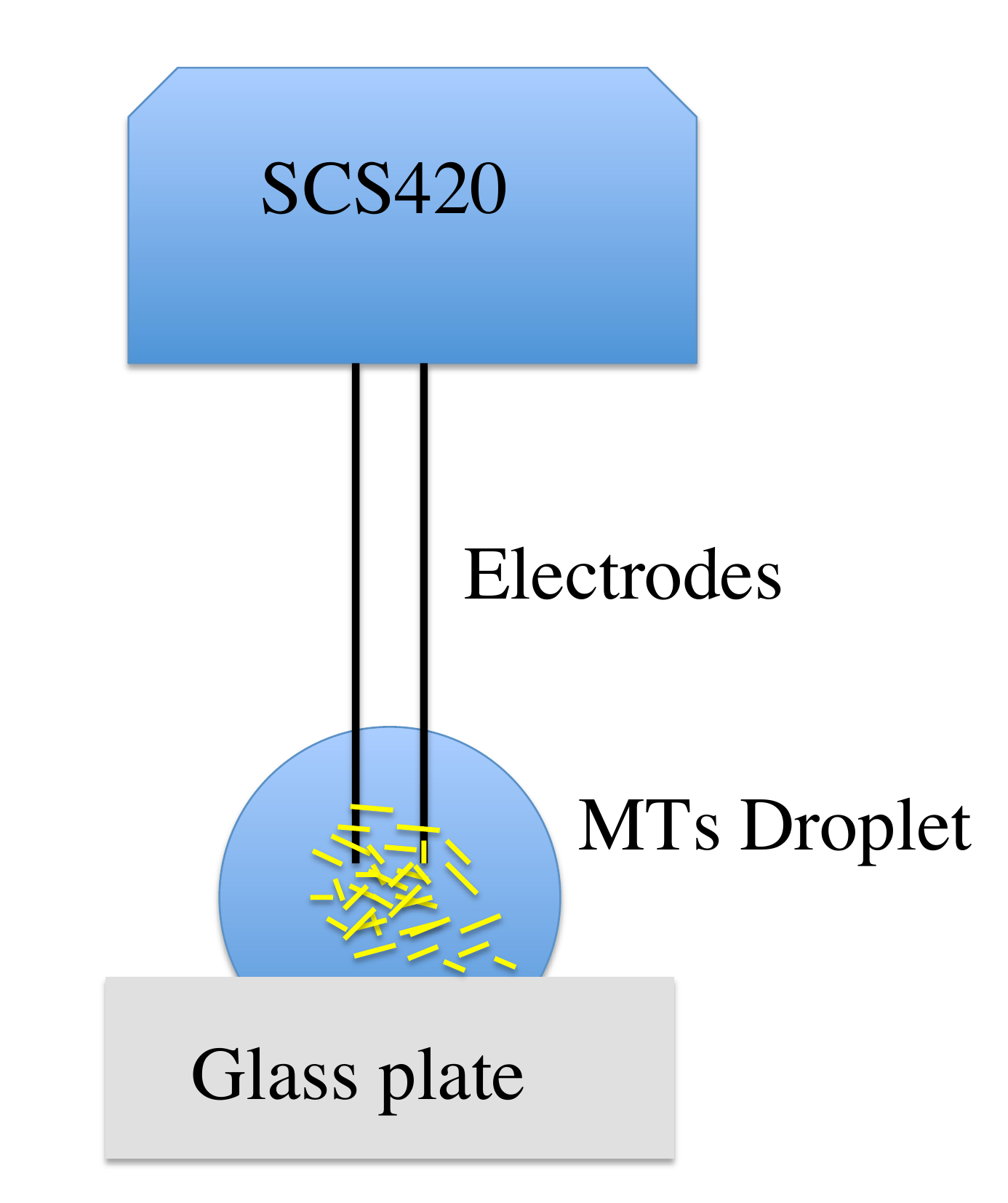}}
    \subfigure[]{\includegraphics[scale=0.1]{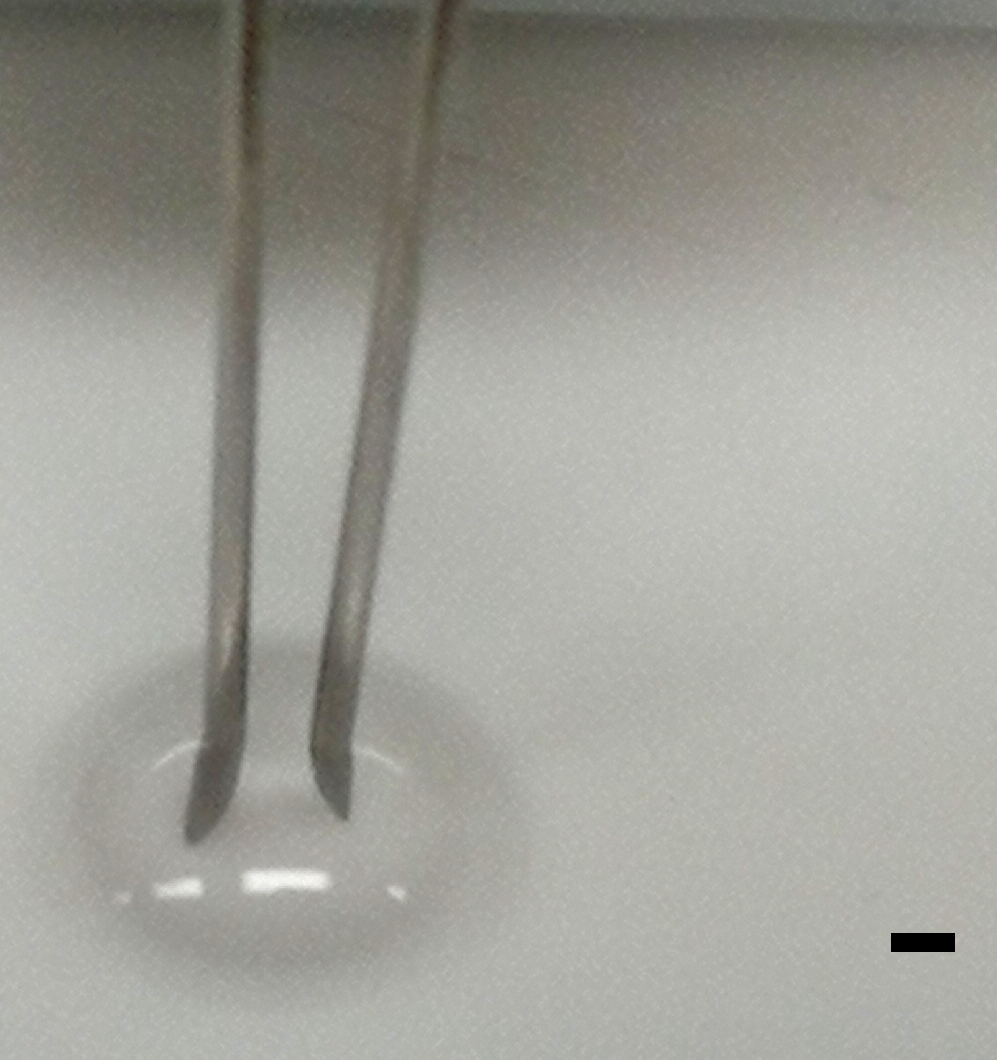}}
    \caption{Scheme of the experiments (a) and a photo of MT droplet with electrodes inserted, bottom right: scale bar 2 mm (b).}
    \label{fig:scheme}
\end{figure}

Electrical stimulation and recording was performed via needle electrodes composed of stainless steel coated by iridium, with twisted cable pairs (Spes Medica Srl, Italy). These electrodes were inserted \SI{1}{\milli\metre} deep into the MT solution droplet, and the distance between them kept at \SI{2}{\milli\metre} (Fig.~\ref{fig:scheme}).  Fresh electrodes were used between each droplet, to prevent the effects of droplet evaporation and MT coating of the electrodes. The electronic characterisation was assessed using a Keithley Semiconductor Characterisation System SCS4200 with triaxial cables and preamplifiers. DC characterisation was conducted in the range $[-1.2, +1.2]$~V. This limit was chosen to avoid the electrolysis of water at \SI{1.23}{\V} and oxidation of Taxol~\cite{Gowda2014}. Alternated current (AC) measurements have been taken in the range from 1~kHz to 10~MHz, typically with a signal amplitude of 10~mV root-mean-square (RMS). By adding a direct current (DC) bias and sweeping it between two saturation values (-0.5~V and +0.5~V), impedance is measured at a fixed frequency with a small signal over-imposed (10~mV RMS).

Pulse Train (PT) measurements featured a high speed pulse generator with internal reference and feedback system that is capable of measuring its own output, plus two independent high speed voltage and current units connected to the device. By simultaneously monitoring the voltage and current at each of the electrodes, the effective resistive response is measured under the application of multiple unipolar pulses. The pulse is rectangular, has an amplitude of 0.5~V and a duration of 1~ms, selected to be comparable to biologic action potentials. The electrical response is recorded for a further \SI{200}{\micro\s}, enabling us to trace the decay of the sample until rest condition (0~V). During each measurement the sample was submitted to a train of 50 pulses separated by 4~ms idle time and the resistive response was recorded and internally averaged; this procedure was repeated 10 times to trace any eventual drift and again averaged to compute the associated standard deviation.

Microscopy was conducted on stabilised MT samples as follows. Samples of prepared MT were fixed in 1\% glutaraldehyde in RB. Fluorescence microscopy was performed on a Zeiss Axiovert 200M inverted light microscope using a $\times 100$ oil immersion lens, following staining with 1\% 4,6-diamidino-2-phenylindole (DAPI) in DMSO; although best known as a nucleic acid stain, DAPI was here used for its affinity for tubulin associated in MTs \cite{Bonne1985}. \SI{25}{\micro\liter} samples were sandwiched between two \SI{1.1}{\milli\meter} glass coverslips and excited with a CoolLed PE100 LED source. Images were collected using a Thorlabs 340M scientific CCD camera (Thorlabs, USA) and associated software. Samples were also observed using transmission electron microscopy (TEM) by adding \SI{2.5}{\micro\liter} of MT solution to a 400 square formvar-coated copper grid for 10 minutes, before adding \SI{25}{\micro\liter} of 1\% uranyl acetate solution, for 30 seconds, after which the solution was blotted. The grids were left to dry, before being imaged in a Philips CM-10 TEM with a Gatan camera system.

\section{Results}
\label{results}

\begin{figure}
    \centering
    \subfigure[]{\includegraphics[width=0.7\textwidth]{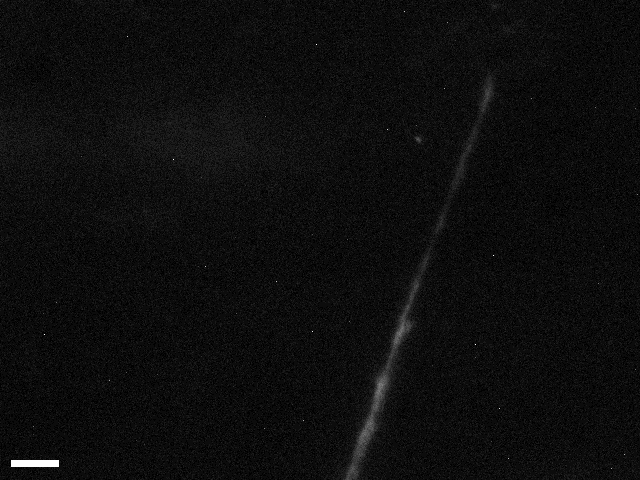}}
    \subfigure[]{\includegraphics[width=0.7\textwidth]{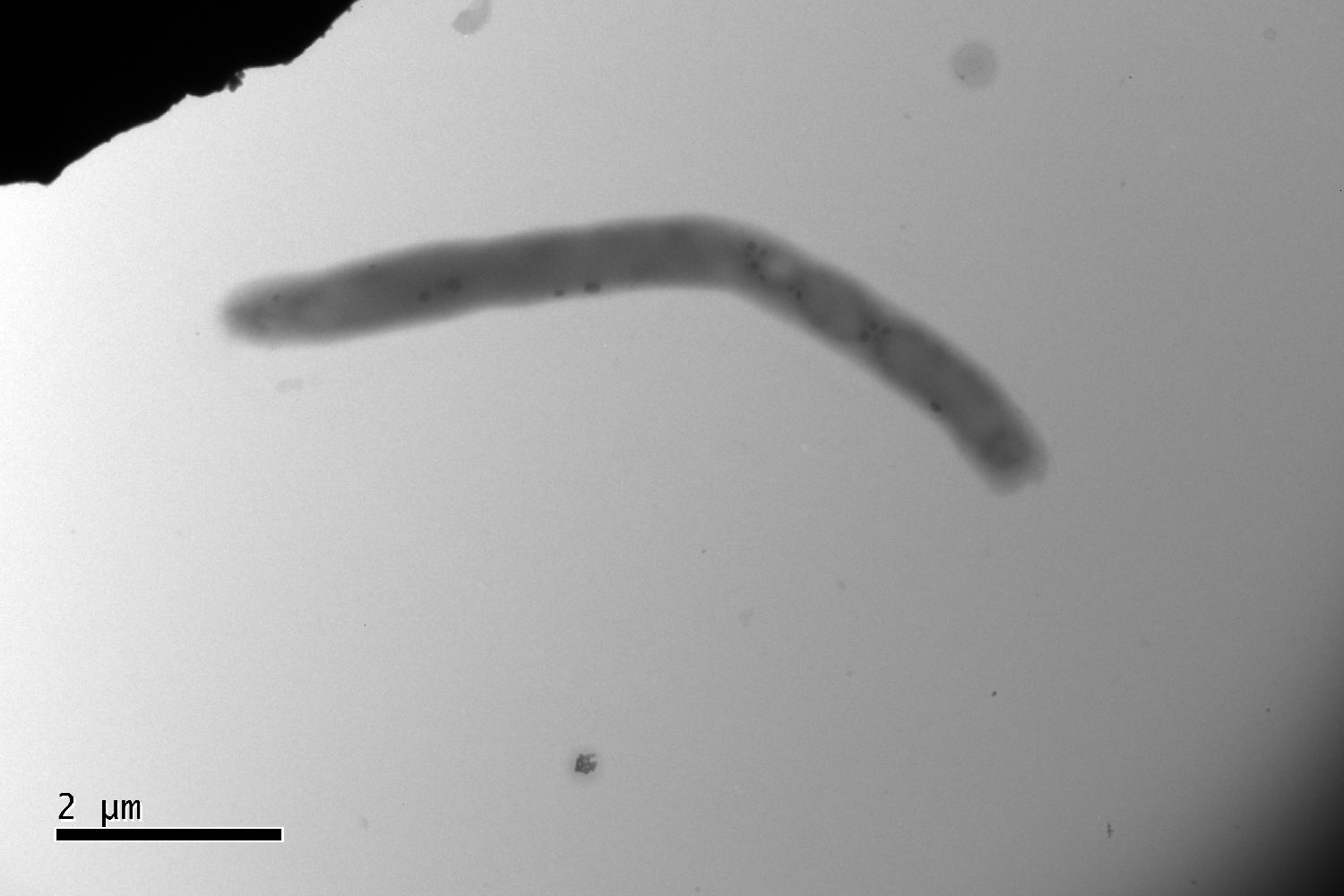}}
    \caption{Microscopic appearance of MT samples. (a) Fluorescence micrograph to show a single stabilised MT. Scale bar \SI{1.0}{\micro\meter}. (b) TEM showing a rod-shaped helical arrangement, containing electron-dense objects (dark circles).}
    \label{fig-microscopy}
\end{figure}

MTs were observed to exist as rod-shaped objects, as expected, via fluorescence microscopy, but were typically longer than their manufacturer-stated length when stabilised with Taxol, at over \SI{10}{\micro\meter} (Fig.~\ref{fig-microscopy})a. No structures correlating with the expected appearance of MT were observed via TEM, but rod-shaped objects that appeared to contain bundles of helically-arranged strings, measuring sub \SI{100}{\nano\meter} and containing multiple electron-dense objects, were observed (Fig.~\ref{fig-microscopy}b).

\FloatBarrier
\begin{finding}
MT droplet acts as an electrical switch. 
\end{finding}

\begin{figure}[!tbp]
    \centering
    \includegraphics[width=0.65\textwidth]{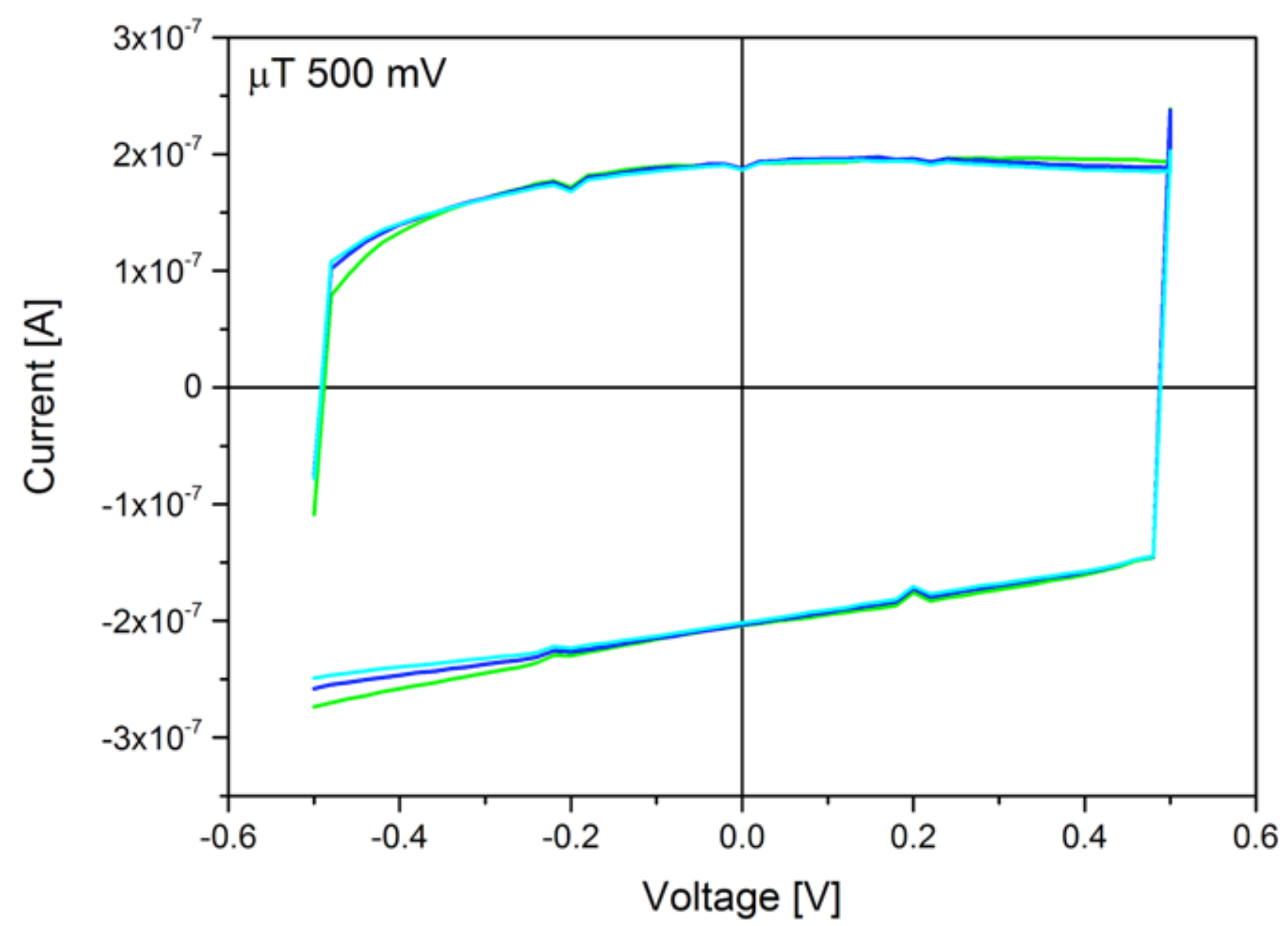}
    \caption{MT IV curve cycles}
    \label{fig:MTSwitch}
\end{figure}

 This was demonstrated by assessing DC properties of stabilised MT (stabilised as described in Sect.~\ref{setup}). Figure~\ref{fig:MTSwitch} shows three IV cycles measured in a range of 0.5~V.  The cycle shows a substantial hysteresis and a maximum current of about 200~nA flowing through the sample, well above the DIW contribution which is one order of magnitude lower. Difference between the upcycle and the downcycle is particularly evident close to saturation: when the voltage is positive, in one case (quadrant IV) we have an associated differential resistance of 
 8.60~M\textOmega ~while in the other case (quadrant I) we have an associated (negative) differential resistance of   -100~M\textOmega. Such negative differential resistance is usually associated to charge transfer phenomena between molecules close enough to enable electrostatic interaction to play a role~\cite{chiolerioporro}.
 
\FloatBarrier
 \begin{finding}
 Stimulation of MT droplet with a positive impulse increases resistance of the droplets and causes oscillations of the droplet’s resistance at the elevated level.
 \end{finding}

\begin{figure}[!tbp]
    \centering
\subfigure[]{\includegraphics[width=0.49\textwidth]{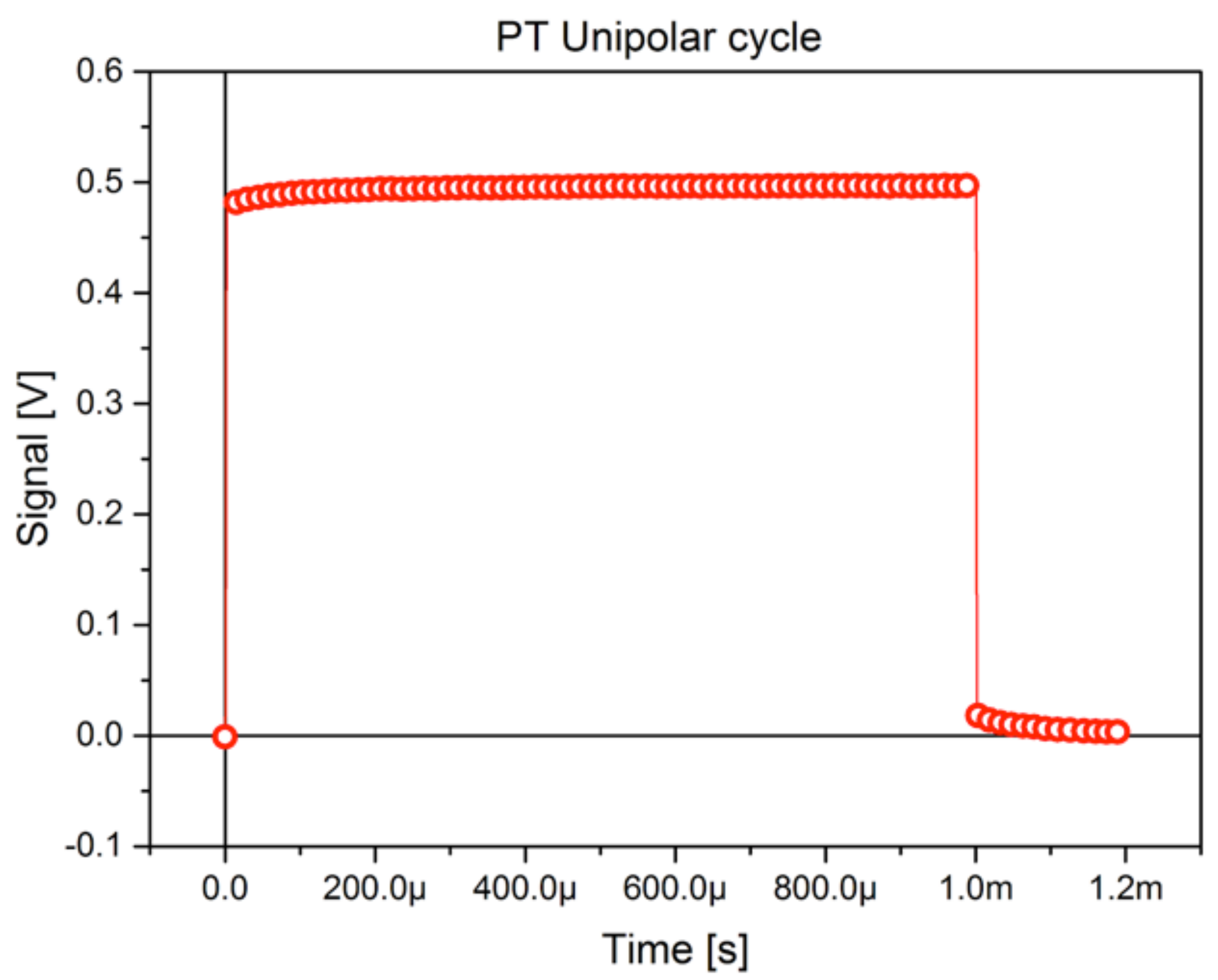}\label{pulse}}
\subfigure[]{\includegraphics[width=0.49\textwidth]{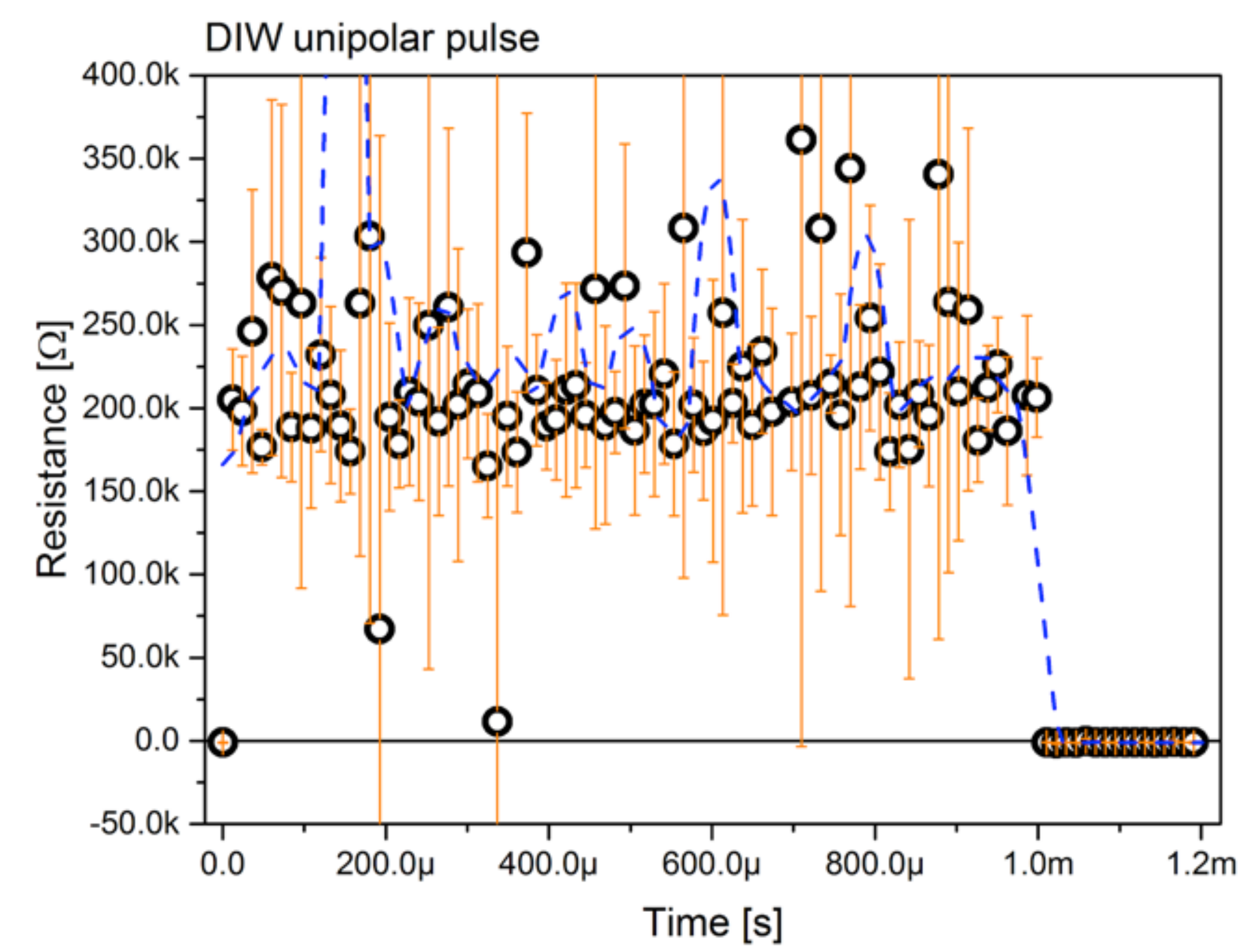}\label{diw_pulse}}
    \caption{PT stimulation of a DIW droplet.
    (a)~Unipolar pulse submitted to samples in the PT characterisation. 
    (b)~DIW response to the same (dashed line: adjacent averaging over 50 points). Ones point every 20 shown for clarity.
    }
\end{figure}

 DIW response to a unipolar pulse (Fig.~\ref{pulse}) is shown in Fig.~\ref{diw_pulse}, evidencing a noisy and rather flat resistance, with no distortion induced by high order harmonics.
 
 \begin{figure}[!tbp]
    \centering
\subfigure[]{\includegraphics[width=0.75\textwidth]{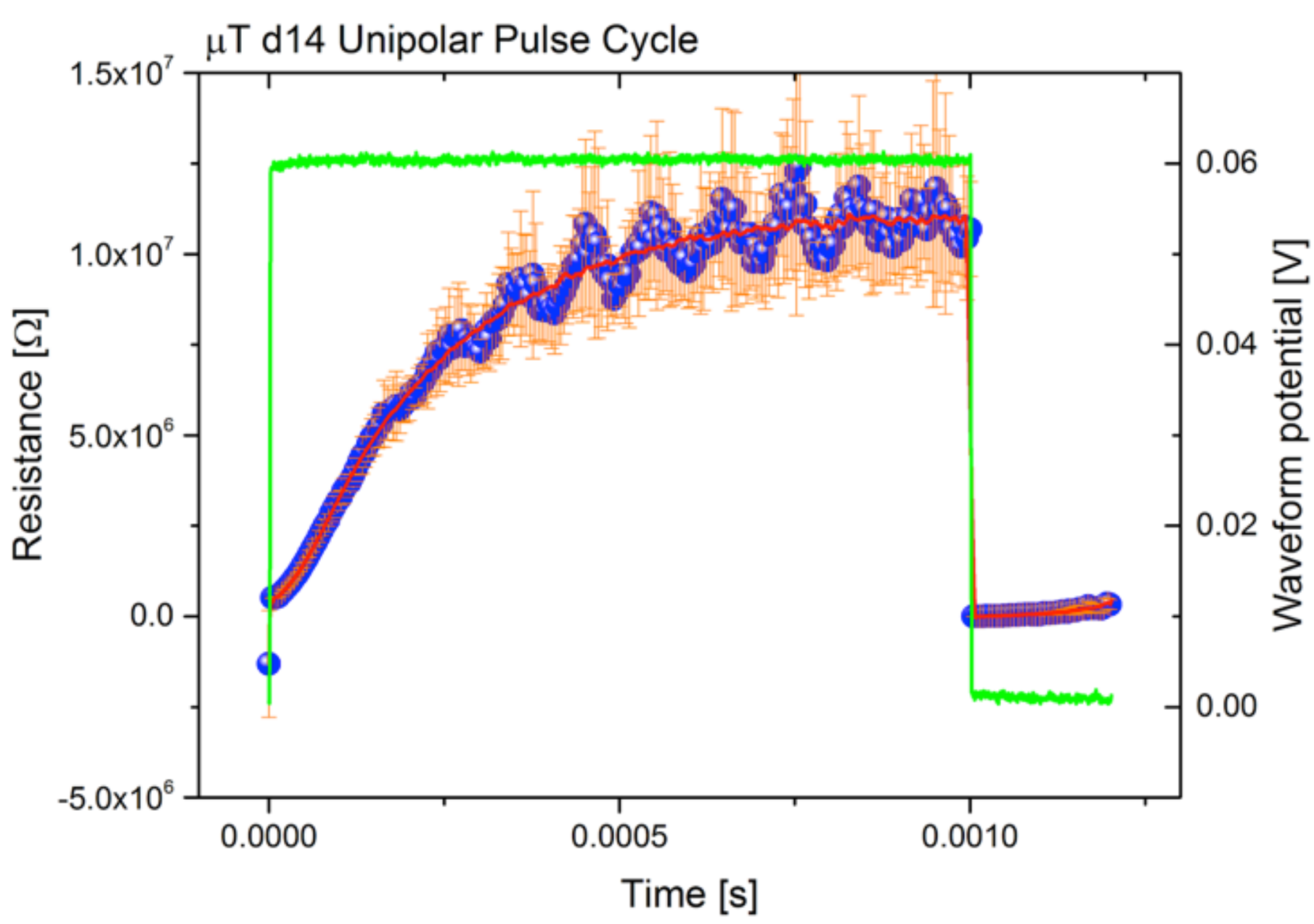}\label{mtpulse1}}
\subfigure[]{\includegraphics[width=0.75\textwidth]{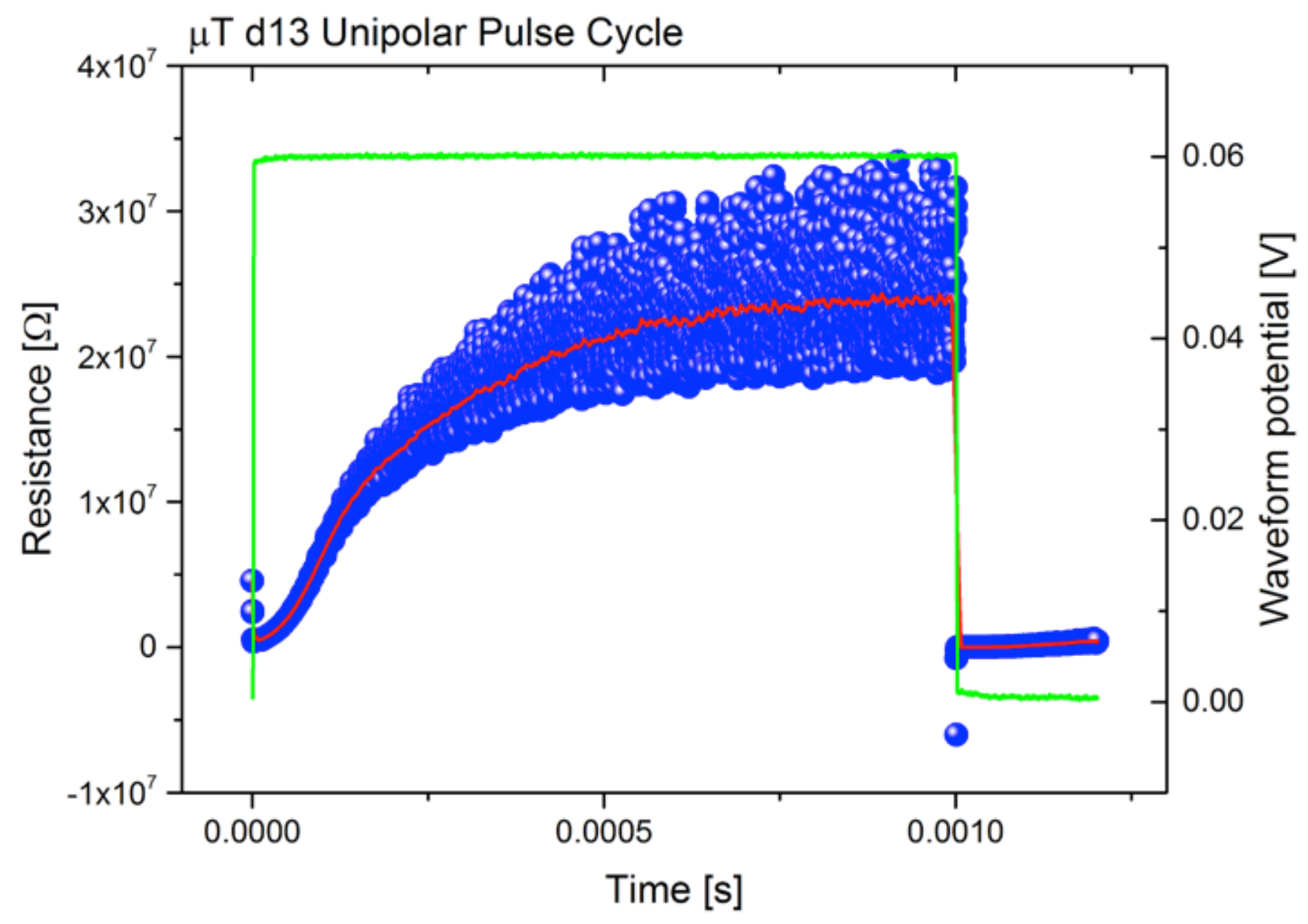}\label{mtpulse2}}
\caption{PT unipolar pulse positive (a), and negative (b), of a MT droplet. Only one point every 15 shown for clarity (blue). Average over six experimental curves superimposed to a smoothing over 51 adjacent points (red). Waveform potential shown in green.
(a)~Dynamics of the droplet resistance in a response to a positive impulse. 
(b)~Dynamics of the droplet resistance in a response to a negative impulse.
    }
\end{figure}

 The rest resistance value is about 950~\textOmega, that we may take as reference for further considerations on other materials. Figures \ref{mtpulse1} and \ref{mtpulse2} shows a comparison between the unipolar pulse response depending on pulse polarity. The positive pulse results (Fig.~\ref{mtpulse1}) in a quite sharp response, plus an oscillating noise on the top profile. The negative pulse results in a very noisy response, rather chaotic (Fig.~\ref{mtpulse2}).

 \begin{figure}[!tbp]
    \centering
\subfigure[]{\includegraphics[width=0.75\textwidth]{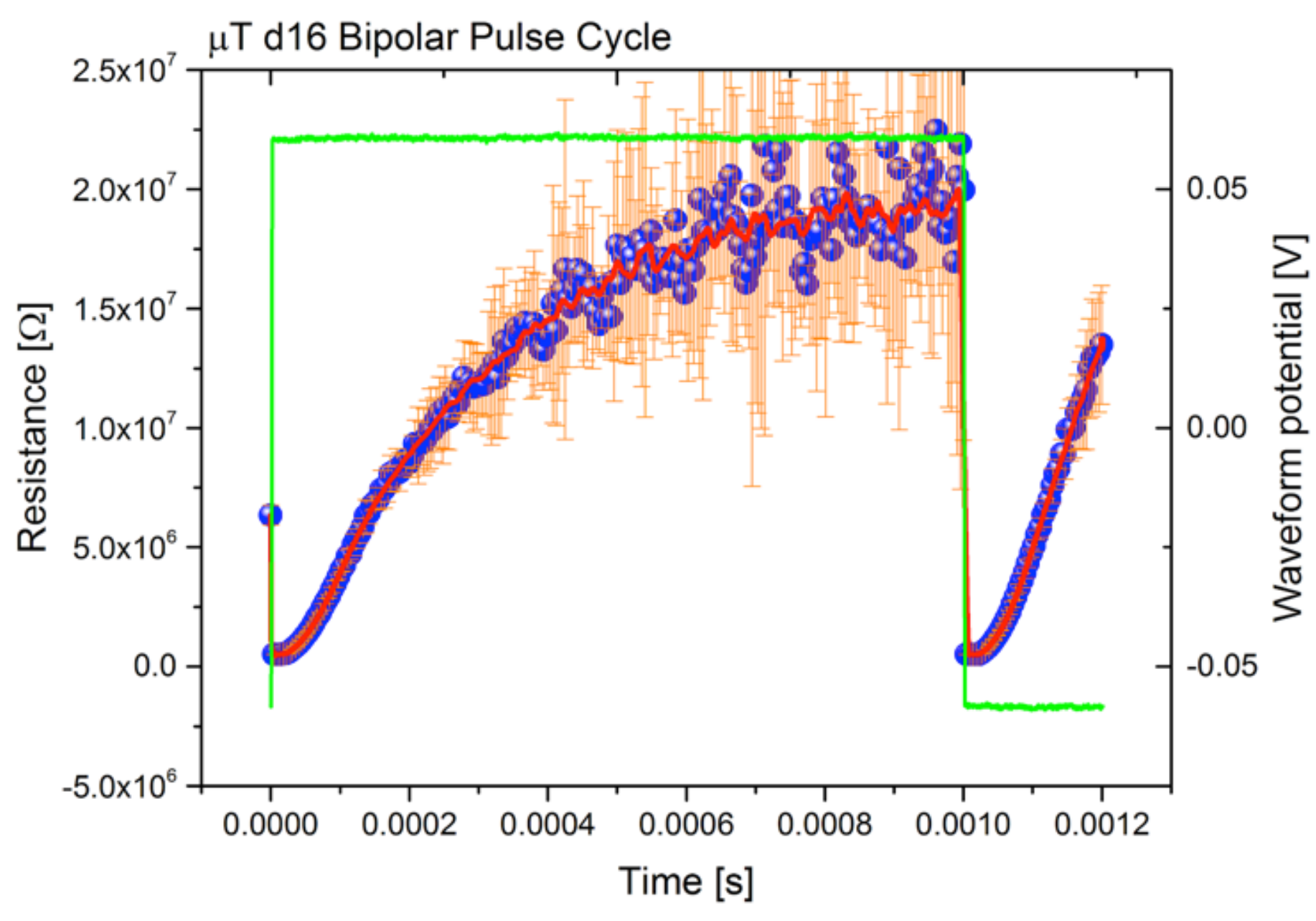}\label{Positivepulse100cycles}}
\subfigure[]{\includegraphics[width=0.75\textwidth]{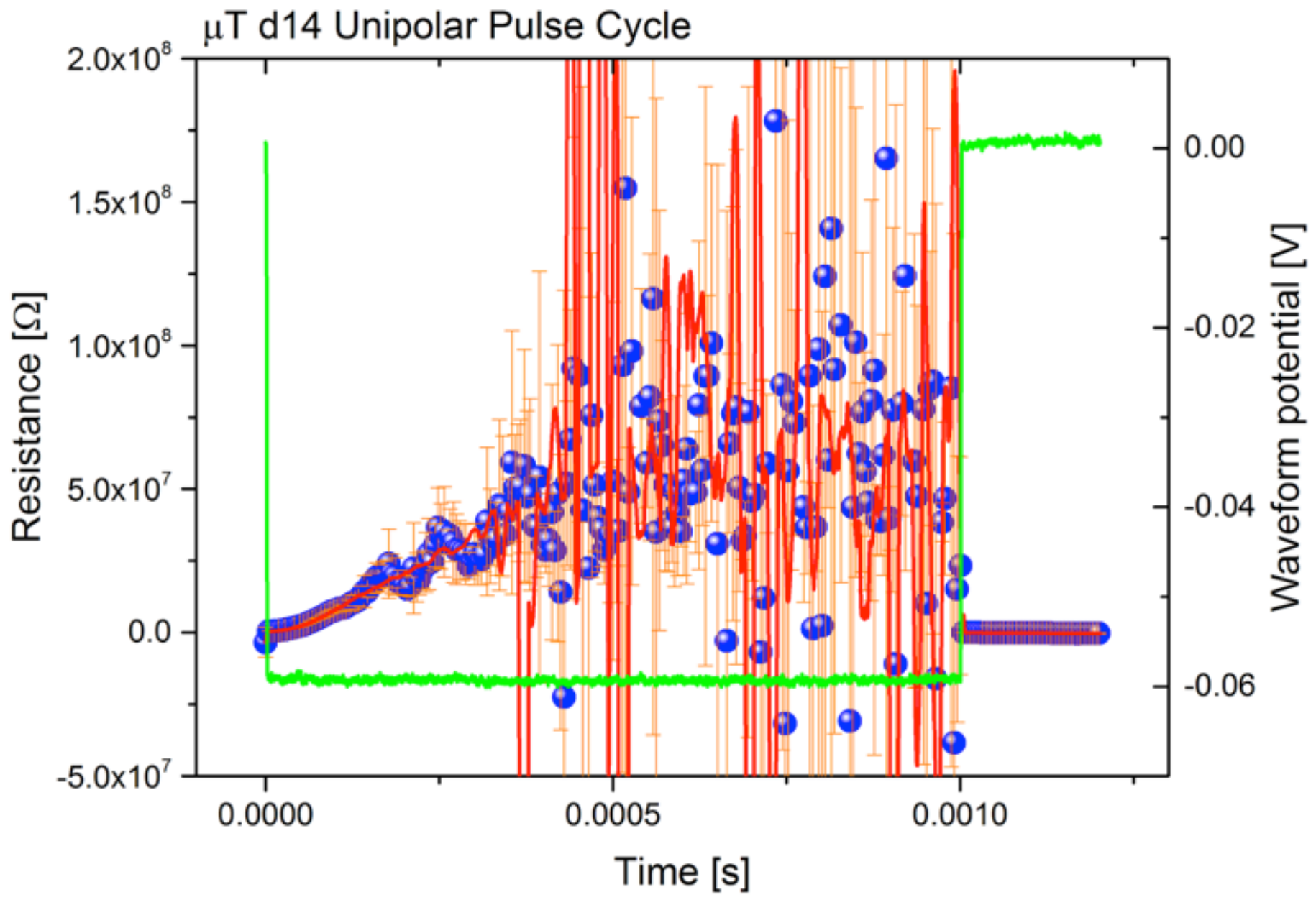}\label{Bipolarpulse100cycles}}
    \caption{Dynamics of the MT droplet resistance in a response to 
    (a)~PT unipolar positive pulse after internal integration over 100 cycles, 
    (b)~PT bipolar. 
    Only one point every 15 shown for clarity (blue).  Waveform potential shown in green.
    }
\end{figure}

 Oscillations of the resistance are not pronounced on the 
 the plot of responses to 100 positive impulses (Fig.~\ref{Positivepulse100cycles}). The integrated responses to a bipolar stimulation does not show any coherent increase in resistance but a wide range of resistance values from \SI{0}{\ohm} to nearly \SI{2e8}{\ohm}.
 

\section{Discussions}
\label{discussion}

We demonstrated that MT droplets can show resistive switching in a limited range of voltages and trigger electronic oscillation under proper electrical stimulus, thus adding new members to a growing family of tubulin electronics~\cite{tuszynski2016microtubule,tuszynski2016microtubule,kalra2019capacitive,gharooni2019bioelectronics}. Future works could be focused on cascading tubulin electronic elements into purposeful electrical analog computing circuits. 

Let us discuss potential mechanisms of the phenomena observed. 
What happens in the droplet? The Taxol molecule binds to each heterodimer in the MT. This binding makes the MT more rigid and helps prevent dissociation of the MT into individual heterodimers~\cite{xiao2006insights}. There are indications that Taxol-stabilised MTs self-organise into bundles of co-aligned tubes~\cite{somers1990kinetics,andreu1992low}. Our TEM results on dehydrated Taxol-stabilised MT samples (Fig. \ref{fig-microscopy}b) would appear to at least partially support this hypothesis. The observed increase in the resistance of MT droplets could be due to  the MTs self-organising into arrays of parallel tubes oriented along the lines of electro-magnetic field~\cite{vassilev1982parallel,brown1999review,glade2005brief}. Let us evaluate a volume of MT networks used in our experiments. 
The radius of a MT is 12.5~nm = $1.25^{-6}$~cm. The length of the MT in the reconstructed solution, as per \cite{tubulinedatasheet}, can be taken as 2~\textmu m = $2^{-4}$~cm.  Thus the volume of one MT is \SI{9.82e-16}{\centi\m\cubed}. The total volume of all MT in each droplet, based on their concentration, is $(1.66 \times 10^{10}) \times (9.82 \times 10^{-16})$ = \SI{1.63e-5}{\centi\m\cubed}. That is just $1/613^{th}$ of the droplet volume is occupied by MT, yet, this amount is sufficient to exhibit non-trivial electrical properties in reason of their percolation threshold and their extremely elongated shape~\cite{castellino}. Note that whilst some MTs longer than \SI{2}{\micro\m} were detected, the concentration and quantity of tubulin in the droplet is fixed and finite, therefore the total occupied space remains unchanged.

\begin{figure}[!tbp]
    \centering
\subfigure[]{\includegraphics[scale=0.55]{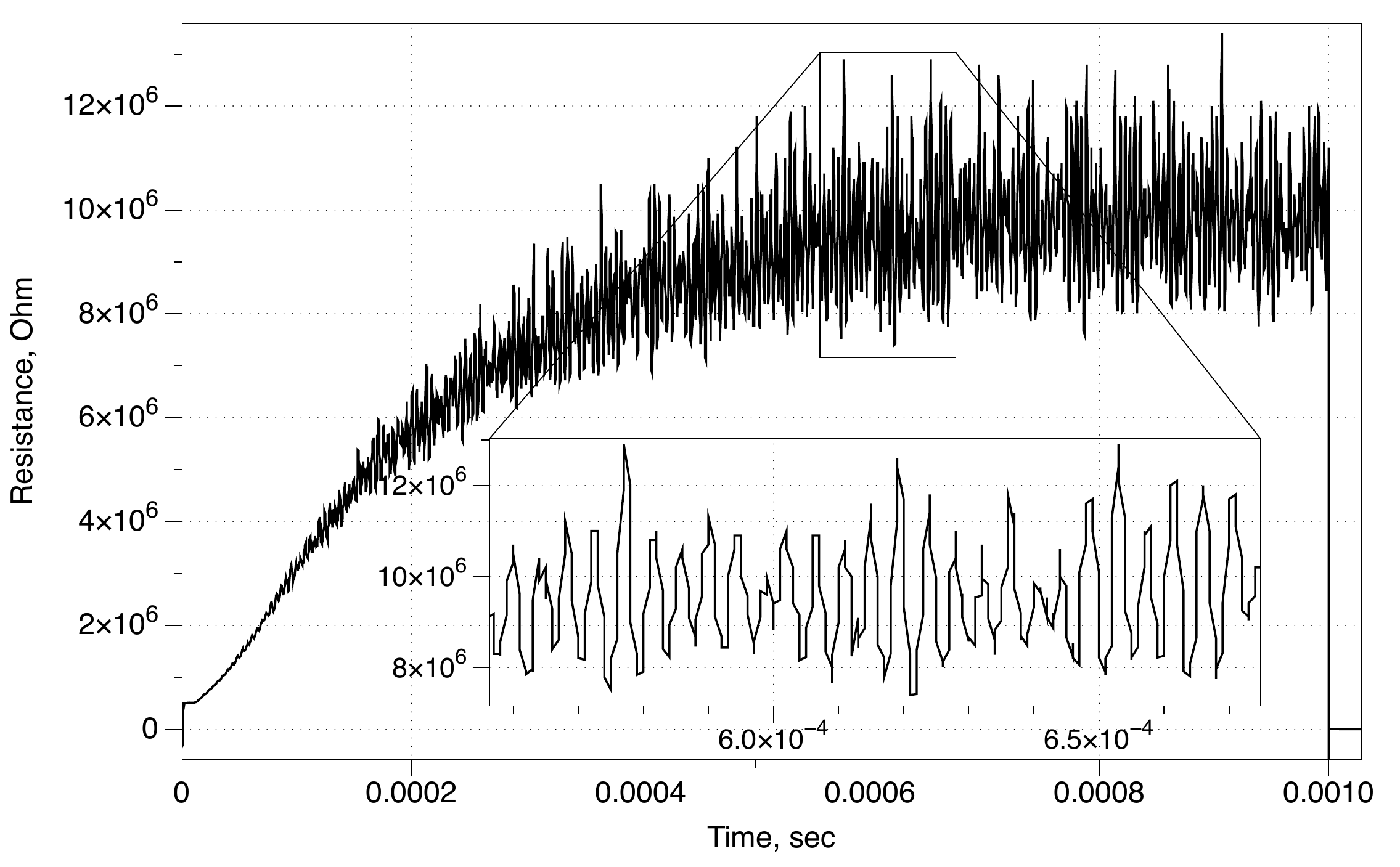}\label{ResistanceOscillation}}
\subfigure[]{\includegraphics[scale=0.55]{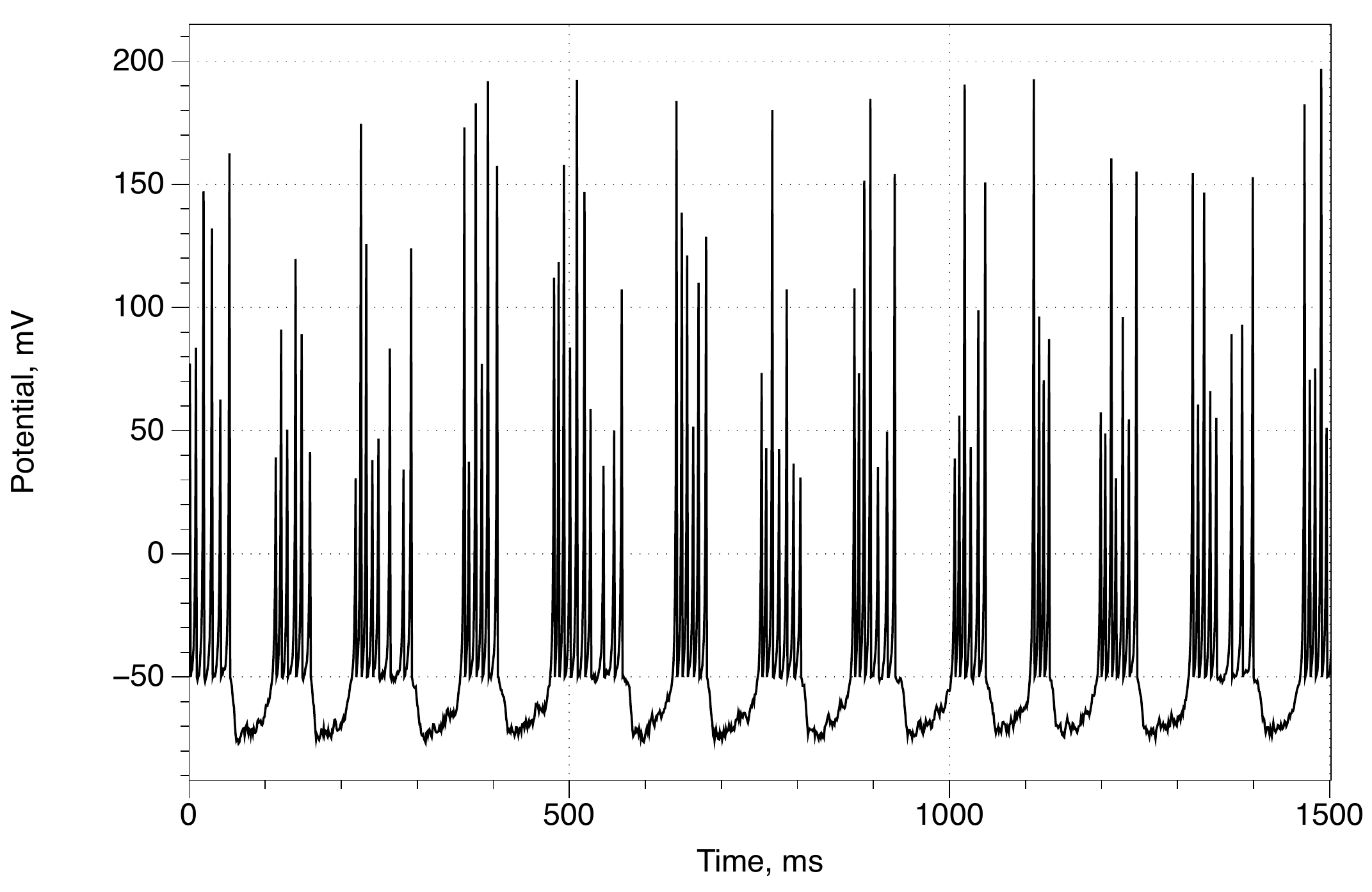}\label{IziOsc}}
    \caption{(a)~Oscillation of resistance.
    (b)~Oscillation of a membrane potential of a computer model of chattering Izikevich neuron.}
    \label{fig:oscillations}
\end{figure}

The MT droplets exhibit oscillations of the resistance under applied positive voltage (Figs.~\ref{mtpulse1} and ~\ref{mtpulse2} and  Fig.~\ref{ResistanceOscillation}). An average period of a resistance spike is 24~ns (2.4$\times 10^{-5}$), standard deviation 0.4~ns. The oscillations can be caused by electro-mechanical vibrations of the MT~\cite{havelka2014multi,li2019electromechanical}, shuttling of conformation changes~\cite{hameroff2002conduction}, orientation transitions of dipole moments~\cite{tuszynski1995ferroelectric,brown1997dipole,cifra2010electric}, and ionic waves~\cite{tuszynski2004ionic,priel2006ionic,sataric2009nonlinear}. The oscillations show a visual resemblance to spiking of a chattering neuron in Izhikevich model~\cite{izhikevich2003simple} (Fig.~\ref{IziOsc}), yet the oscillations time scale is different by several orders of magnitude. Those faster oscillations could be the result of the much higher stiffness of MT bundles in Taxol-stabilised solutions. Further work is required to evaluate the possibility of integration of nano-seconds scale spiking in MT network into milli-seconds range spiking of neurons to understand better how exactly the MT oscillation can contribute to neuronal activity and information processing~\cite{craddock2010critical,salari2011photonic}.

\section{Conclusions}
\label{conc}

We have demonstrated some of the  electrical properties of aqueous Taxol-stabilised MTs. By applying a cycling DC potential and monitoring the current, we have indicated the electrical switch capabilities of the system, inherent in its large hysteresis. We have also revealed oscillations of the systems resistances, when exposed to the PT experiments. These oscillations could have a number of sources, and require further investigation. The findings add to the body of knowledge for MT-based systems, permitting further work in fields such as sensing, computing, and data storage.


\section*{Acknowledgements}

 AC acknowledges Fondazione Istituto Italiano di Tecnologia for funding. Measurements were taken at the University of the West of England, Bristol, UK, while AC was visiting. AA and TCD were partly supported by by the EPSRC with grant \mbox{EP/P016677/1}.

RM acknowledges Dr David Patton and Mrs Sue Hula of UWE Bristol for their invaluable expertise with the electron microscopical aspects of this work.

\section*{References}

\end{document}